\documentclass[aps,prl,superscriptaddress,twocolumn,showpacs,preprintnumbers,amsmath,amssymb]{revtex4-1}
\usepackage{graphicx}
\usepackage{dcolumn}
\usepackage{bm}
\usepackage{upgreek}
\usepackage{placeins}

\begin{document}

\title{Low anisotropy of the upper critical field in a strongly anisotropic layered
cuprate Bi$_{2.15}$Sr$_{1.9}$CuO$_{6+\delta}$: Evidence for a
paramagnetically limited superconductivity.}

\author{S. O. Katterwe$^\dagger$}
\author{Th. Jacobs}
\affiliation{Department of Physics, Stockholm University, AlbaNova
University Center, SE-10691 Stockholm, Sweden}

\author{A. Maljuk}
\affiliation{Leibniz Institute for Solid State and Materials
Research IFW Dresden, Helmholtzstr. 20, D-01171 Dresden, Germany}

\author{V. M. Krasnov$^1$}
\email[E-mail: ]{Vladimir.Krasnov@fysik.su.se}

\date{\today }

\begin{abstract}

We study angular-dependent magnetoresistance in a low $T_c$
layered cuprate Bi$_{2.15}$Sr$_{1.9}$CuO$_{6+\delta}$. The low
$T_c \sim 4$~K allows complete suppression of superconductivity by
modest magnetic fields and facilitate accurate analysis of the
upper critical field $H_{c2}$. We observe an universal exponential
decay of fluctuation conductivity in a broad range of temperatures
above $T_c$ and propose a new method for extraction of $H_{c2}(T)$
from the scaling analysis of the fluctuation conductivity at
$T>T_c$. Our main result is observation of a surprisingly low
$H_{c2}$ anisotropy $\sim 2$, which is much smaller than the
effective mass anisotropy of the material $\sim 300$. We show that
the anisotropy is decreasing with increasing field and saturates
at a small value when the field reaches the paramagnetic limit. We
argue that the dramatic discrepancy of high field and low field
anisotropies is a clear evidence for paramagnetically limited
superconductivity.

\end{abstract}

\maketitle

\section{Introduction}

The upper critical field $H_{c2}$ is one of the key parameters of
type-II superconductors \cite{SaintJames}. It is particularly
important for understanding unconventional superconductivity
\cite{Gurevich2007,Prozorov2012}. However, estimation of $H_{c2}$
for high-temperature superconductors is a notoriously difficult
task. The high $T_c$ leads to an extended region of thermally
activated flux-flow. The complex physics of anisotropic pinning
and melting of the vortex lattice \cite{Blatter} makes it hard, if
at all possible \cite{Vedeneev1999}, to confidently obtain
$H_{c2}$ from flux-flow characteristics at $T<T_c(H)$.

The high $T_c$ in combination with a strong coupling leads to a
large superconducting energy gap $\Delta \sim 20-50$ meV
\cite{KrTemp,Hoffman2002,Fischer2007,SecondOrder,ARPESreview,Ideta_ARPES}
and $H_{c2}(0) \sim 10^2$ T
\cite{Obrian2000,Sekitani2004,MR,Vedeneev2006,Li2007,Ramshaw2012,Taillifer2012,Grissonnanche_2014}.
Such strong fields may alter the ground state of the material. For
cuprates and pnictides the parent state is antiferromagnetic. It
has been demonstrated that relatively weak fields can induce a
canted ferromagnetic order in strongly underdoped cuprates
\cite{Ando_2003}.
Furthermore, the normal state of underdoped cuprates is
characterised by the presence of the pseudogap (PG), which
probably represents a charge/spin or orbital density wave order
coexisting and competing with superconductivity
\cite{Tallon2001,SecondOrder,ARPESreview,Vishik2012,Onufrieva2012,d_density,Varma2002,Weber2009,Orenstein2013,Gabovich2013}.
Suppression of superconductivity by magnetic field may enhance the
competing PG, as follows from observation of a charge density wave
in a vortex core \cite{Hoffman2002}. But even stronger magnetic
fields of several hundred tesla suppress the PG
\cite{KElbaum2005,Jacobs2012}. Thus, both superconducting and
normal state properties are affected by strong magnetic fields and
separation of the two contributions is highly non-trivial and
controversial. Disentanglement of superconducting and PG
characteristics is difficult even above $T_c$ due to presence of
profound superconducting fluctuations
\cite{Dubroka_2011,Kaminski_NP2011,Varlamov}. Therefore, principal
new questions, which do not appear for low-$T_c$ superconductors,
are to what extent $H\sim H_{c2}$ alters the abnormal normal state
of high-$T_c$ superconductors and how to define the
non-superconducting background in measured characteristics.

For many unconventional superconductors the measured $H_{c2}$
exceeds the paramagnetic limit of the BCS theory
\cite{SaintJames,Clogston}. This has been reported for organic
\cite{Zuo2000,Singleton2000,Lee2000}, cuprate
\cite{Vedeneev2006,Li2007}, pnictide
\cite{Cho2011,Khim2011,Burger2013} and heavy fermion
\cite{Radovan2003,Kakuyanagi2005,Matsuda2007} superconductors.
Such an overshooting is an important hint in a long standing
search for exotic spin-triplet and
Fulde-Ferrell-Larkin-Ovchinnikov states (for review see e.g.
Ref.\cite{Matsuda2007}). Yet, the overshooting is not a proof of
unconventional pairing because the paramagnetic limit is rather
flexible. It is increasing in the presence of spin-orbit
interaction \cite{SaintJames} and in the two-dimensional (2D) case
and is lifted in the one-dimensional (1D) case
\cite{Buzdin1996,Matsuda2007}. Unconventional superconductors are
usually anisotropic. Some of them have quasi-2D, or possibly even
quasi-1D structure.
Many have a significant spin-orbit interaction between localized
spins and itinerant charge carriers. Consequently, one needs a
more robust criterion for the paramagnetically (un)limited
superconductivity in search for exotic states of matter.

Here we investigate the anisotropy of $H_{c2}$ in a strongly
anisotropic layered Bi$_{2.15}$Sr$_{1.9}$CuO$_{6+\delta}$
(Bi-2201) cuprate with a low $T_c \sim 4$ K. The low $T_c$ and the
associated large disparity of superconducting and pseudogap scales
\cite{Jacobs2012} allow simple and accurate estimation of $H_{c2}$
without complications typical for high-$T_c$ cuprates. We present
a detailed analysis of angular dependence of in-plane and
out-of-plane magnetoresistances (MR) and demonstrate that they
exhibit very different behavior. We observe an universal
approximately exponential decay of the in-plane fluctuation
para-conductivity above $T_c$ and propose a method for extraction
of $H_{c2}(T)$ from a new type of a scaling analysis of
fluctuations at $T>T_c$. It obviates the complexity of the
flux-flow phenomena and allows unambiguous extraction of
$H_{c2}(T)$. Remarkably, we obtained that the anisotropy of the
upper critical field
$H_{c2}^{\parallel}/H_{c2}^{\perp}(T\rightarrow 0) \simeq 2$ is
much smaller than the anisotropy of the effective mass $\gamma_m
\simeq 300$ \cite{Vinnikov}. This discrepancy clearly indicates
that $H_{c2}^{\parallel}$ parallel to the CuO planes is cut-off by
the paramagnetic limit.

Cuprates have homologous families with different number of CuO
planes per unit cell. Cuprates within the homologous family have
similar carrier concentrations, resistivities, anisotropies and
layeredness, but largely different $T_\text{c}$. For Bi-based
cuprates the three-layer compound
Bi$_{2}$Sr$_{2}$Ca$_2$Cu$_3$O$_{10+\delta}$ (Bi-2223) has a
maximum $T_c$ of $\sim 110$ K, the two-layer compound
Bi$_{2}$Sr$_{2}$CaCu$_2$O$_{8+\delta}$ (Bi-2212) has a $T_c \sim
95$ K and a single-layer compound Bi$_{2}$Sr$_{2}$CuO$_{6+\delta}$
(Bi-2201) has an optimal (with respect to Oxygen doping) $T_c$
that ranges from $\sim 30$ K for Bi/Pb and Sr/La substituted
crystals \cite{Lavrov} to just few K in the pure Bi-2201 compound
\cite{Vedeneev1999,Sonder1989,Maljuk,Luo2014}. According to Ref.
\cite{Maljuk} the stoichiometric Bi-2201 compound is
non-superconducting and a finite $T_c$ appears only in
off-stoichiometric Bi$_{2+x}$Sr$_{2-y}$CuO$_{6+\delta}$ compounds
with $x,y \neq 0$. Thus, the Bi/Sr off-stoichiometry allows fine
tuning of the maximum $T_c$ \cite{Maljuk,Luo2014}.

Development of high magnetic field techniques in recent years has
lead to a significant progress in studies of $H_{c2}$ in
high-$T_c$ superconductors
\cite{Sekitani2004,Vedeneev2006,Taillifer2012,Grissonnanche_2014}.
But the problem of disentanglement of superconducting and PG
magnetic responses remains. It leads to a lack of clear criteria
for extraction of $H_{c2}$ from measurement at $H\sim 100$ T. This
problem is avoided in low-$T_c$ cuprates because the relative
disparity between superconducting and pseudogap scales is
increasing with decreasing $T_c$ \cite{Jacobs2012}. Therefore,
analysis of $H_{c2}$ in low-$T_c$ cuprates should provide an
unambiguous information about the superconducting state, not
affected by interference with the co-existing PG. This is the main
motivation of the present work.

\begin{figure*}[t]
    \centering
    \includegraphics[width=\textwidth]{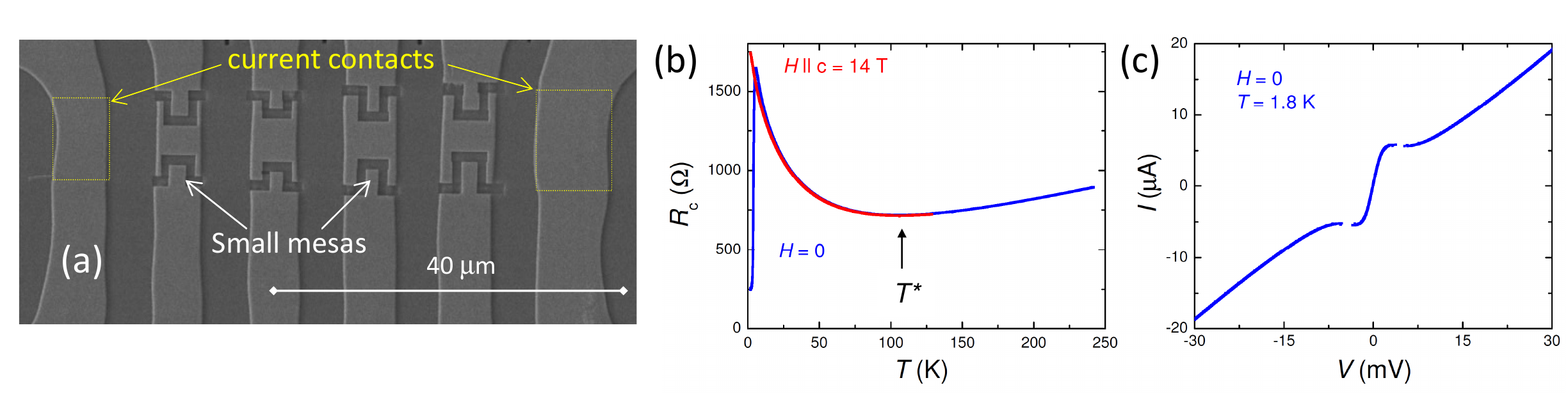}
    \caption{(color online)
(a) Scanning electron microscopy image of the sample OP(4.3). (b)
Temperatures dependence of the $c$-axis resistance at $H=0$ and 14
T. The compound has a low $T_c\simeq 4 K$ and the pseudogap onset
temperature $T^* \simeq 110$ K. (c) Current-voltage characteristic
of a small mesa
at $T=1.8$ K and $H=0$. 
}
    \label{fig:fig1}
\end{figure*}

\section{Experimental}

Studied crystals are parts of one pristine
Bi$_{2.15}$Sr$_{1.9}$CuO$_{6+\delta}$ single crystal with $T_c
\simeq 3.5$ K. Growth and characterization of crystals is
described in Ref.~\cite{Maljuk}. Oxygen doping was consecutively
decreased by soft annealing in vacuum, which does not affect the
crystal quality \cite{Sonder1989}. We present data for a slightly
overdoped (with respect to oxygen content) $T_c \simeq 4.0$ K
[OD(4.0)] and a nearly optimally doped $T_c \simeq 4.3$ K
[OP(4.3)] crystals.

Figure \ref{fig:fig1} (a) shows an image of the studied sample
OP(4.3). The sample consists of ten micron-size mesa structures
(two big and eight small) with attached gold electrodes.
In-plane resistance is measured with a lock-in technique in a
four-probe configuration by sending an ac-current through the left
and right current contacts (big mesas), and measuring the
longitudinal voltage between a pair of small mesas. The $c$-axis
transport is measured in a three-probe configuration by sending a
probe current through one of the small mesas to one of the current
contacts. The voltage is measured with respect to unbiased contact
pad.
Details of sample fabrication and measurement setup can be found
in Ref. \cite{Jacobs2012}.

Fig. \ref{fig:fig1} (b) shows the $c$-axis resistance versus
temperature at $H=0$ and $14$ T along the $c$-axis. It is seen
that $R_c(T)$ exhibits an upturn at $T<T^*\sim 110$ K, indicating
opening of the $c$-axis PG. According to previous studies
\cite{Lavrov,Yurgens_Bi2201,Jacobs2012} such a $T^*$ corresponds
to a near optimally doped (OP) (slightly underdoped) Bi-2201. A
superconducting transition occurs at a much lower $T_c \simeq 4$
K. The $c$-axis field of 14 T completely suppresses the
superconducting transition but does not change significantly the
PG characteristics due to a large disparity of superconducting and
PG scales in this low-$T_c$ compound \cite{Jacobs2012}.

The large $c$-axis resistance $R_c\sim k\Omega$ corresponds to a
non-metallic resistivity $\rho_c \simeq 20~\Omega$cm
\cite{Jacobs2012}, which is much larger than the in-plane
resistivity $\rho_{ab}\simeq 1-4\cdot 10^{-4}~\Omega$cm
\cite{Ando1996}. The anisotropy of resistivity
$\gamma_{R}=\rho_c/\rho_{ab} \sim 10^5$ and the corresponding
effective mass anisotropy $\gamma_m = \sqrt{\gamma_R} \sim 300$ is
very large \cite{Vinnikov}, similar to Bi-2212 \cite{Watanabe1997}
and Bi-2223 \cite{Suzuki_Bi2223} cuprates. This reflects a layered
2D structure of Bi-based cuprates with mobile electrons localized
on atomic CuO planes. The $c$-axis transport is caused by
interlayer tunneling. Below $T_c$ this leads to appearance of an
intrinsic Josephson effect \cite{Kleiner}, observed in all
Bi-based cuprates
\cite{KrTemp,Katterwe2009,SecondOrder,MR,Suzuki_Bi2223}, including
Bi-2201 \cite{Yurgens_Bi2201,MQT_Bi2201,Jacobs2012}. Interlayer
tunneling creates the basis for the intrinsic tunneling
spectroscopy technique
\cite{KrTemp,Suzuki_Bi2223,SecondOrder,MR,Jacobs2012} and
facilitates simultaneous magneto-transport and spectroscopic
measurements, beneficial for analysis of $H_{c2}$ \cite{MR}. Fig.
\ref{fig:fig1} (c) shows the current-voltage $I$-$V$
characteristics of a small mesa at $T=1.8$ K. A detailed analysis
of intrinsic tunneling characteristics of our Bi-2201 crystals can
be found in Ref. \cite{Jacobs2012}. Small area of our mesas allows
investigation of intrinsic tunneling characteristics
\cite{Jacobs2012,SecondOrder,MR,Suzuki_Bi2223,KrTemp} without
significant distortion by self-heating \cite{SecondOrder}.

\section{In-plane and out-of-plain magnetoresistance}

\subsection{A. In-plane magnetoresistance}

\begin{figure*}[t]
    \includegraphics[width=\textwidth]{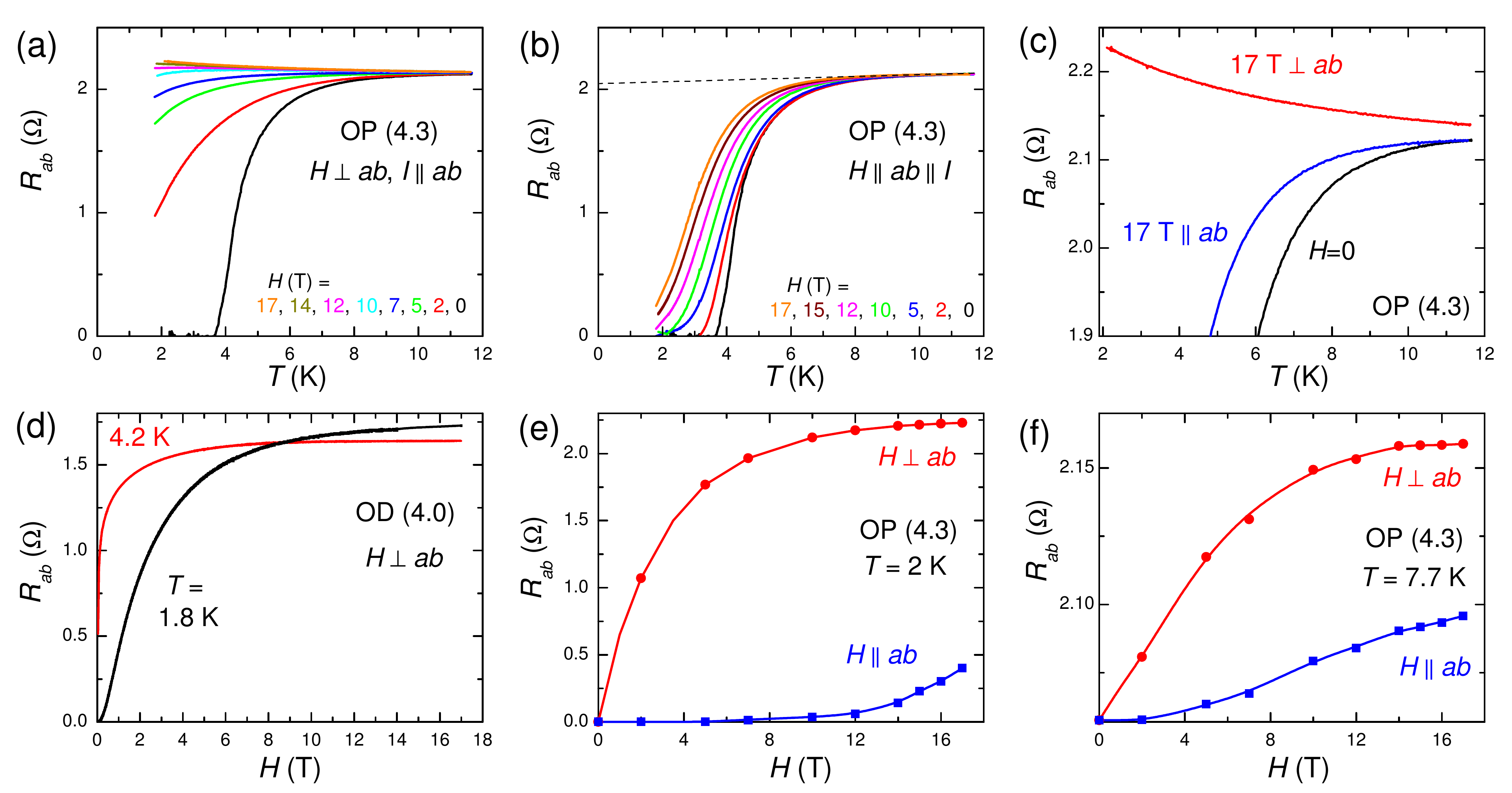}
    \caption{(Color online)
$T$-dependencies of the in-plane resistance at magnetic fields (a)
perpendicular and (b) parallel to the $ab$-planes.
(c) Comparison of the data from (a) and (b) for zero and 17 T. For
$H\parallel ab$ the $R_{ab}(T)$ is shifted to lower temperatures.
For $H\perp ab$ it is completely suppressed and $R_{ab}$ is
shifted upwards, indicating presence of a positive orbital
magnetoresistance in the normal state. (d) MR in a perpendicular
field below and just above $T_c$. Note that the saturation field
$\sim H_{c2}(T)$ is decreasing with $T\rightarrow T_c$. Panels (e)
and (f) show MR for both field orientations (e) below and (f)
above $T_c$. The positive MR at $T>T_c$ is caused both by
suppression of superconducting fluctuations and an additional
orbital normal state MR.}
    \label{fig:fig2}
\end{figure*}

Figures \ref{fig:fig2} (a) and (b) show temperature dependencies
of the in-plane resistance $R_{ab}$ at different magnetic fields
(a) perpendicular and (b) parallel to the $ab$ planes for the
OP(4.3) sample. For $H\perp ab$, $R_{ab}$ reaches the normal state
value $R_n$ already at $H\simeq 10$ T. For $H\parallel ab$ the
field of 17 T still does not completely suppress
superconductivity. The difference is both due to the anisotropy
and due to different contributions from flux-flow and orbital
effects. The Lorentz force density $f_L=(1/c)[J \times B]$, where
$J$ is the transport current density and $B$ is the magnetic
induction, acts both on vortices and mobile charge carriers. In
Fig. \ref{fig:fig2} (a) $H \perp I \parallel ab$ the Lorentz force
is at maximum and effectively drives pancake vortices
\cite{Blatter} along CuO planes. Therefore, $R_{ab}(H\perp ab)$ is
dominated by the flux-flow contribution at $T<T_c(H)$. In case of
Fig. \ref{fig:fig2} (b) $H \parallel ab \parallel I$ there is no
Lorentz force and the flux-flow contribution should be minimal.

Fig. \ref{fig:fig2} (c) represents a detailed comparison of
$R_{ab}(T)$ at $H=0$ and 17 T for the two field orientations. We
notice that the resistive transition at $H
\parallel ab \parallel I$ is simply shifted towards a lower
$T$ due to suppression of $T_c(H)$. On the other hand $R_{ab}$ at
$H \perp ab$ is also shifted upwards, even at $T \gg T_c$. It
indicats that there is an additional positive MR in the normal
state ($\sim 1 \%$ at $H\perp ab = 17$ T). Thus, there are two
different mechanisms of positive in-plane MR. At $T \lesssim T_c$
it is mostly due to suppression of superconductivity. Such MR
saturates at $H \sim H_{c2}$. Fig. \ref{fig:fig2} (d) shows
field-dependence of $R_{ab}(H^{\perp})$ at $T=1.8$ K and at
$T=4.2~{\text K} \sim T_c$. It is seen that saturation of
$R_{ab}(H^{\perp})$ occurs at significantly lower field for $T=4.2
$ K, consistent with reduction of $H_{c2}$ at $T\rightarrow T_c$.
In the normal state $T>T_c$ the tendency is reversed. With
increasing $T$ the saturation field is increasing. This can be
seen from Figs. \ref{fig:fig2} (e) and (f), which show
field-dependence of $R_{ab}$ in perpendicular (circles) and
parallel (squares) magnetic fields at $T=2$ K and $7.7$ K,
respectively. Such behavior can be partly attributed to
superconducting fluctuations, for which the characteristic field
is increasing with $\mid T_c-T \mid$ \cite{Varlamov}. However,
fluctuations do not explain the increment of the saturation value
of $R_n$, which is visible at $T\gg T_c$ and is significant only
for $H \perp ab$, see Fig. \ref{fig:fig2} (c). Consequently, there
is an additional normal state MR, caused by orbiting of mobile
electrons in magnetic field \cite{Ziman}. This leads to a positive
MR with saturation at $\omega_c \tau>1$, where $\omega_c=eB/mc$ is
the cyclotron frequency and $\tau$ is the scattering time. Since
$\tau$ becomes shorter with increasing $T$, the saturation field
is increasing with increasing $T$. Due to the quasi-2D electronic
structure of Bi-2201, the orbital MR should appear only at $H
\perp ab$, consistent with our observation.

\begin{figure*}[t]
    \includegraphics[width=\textwidth]{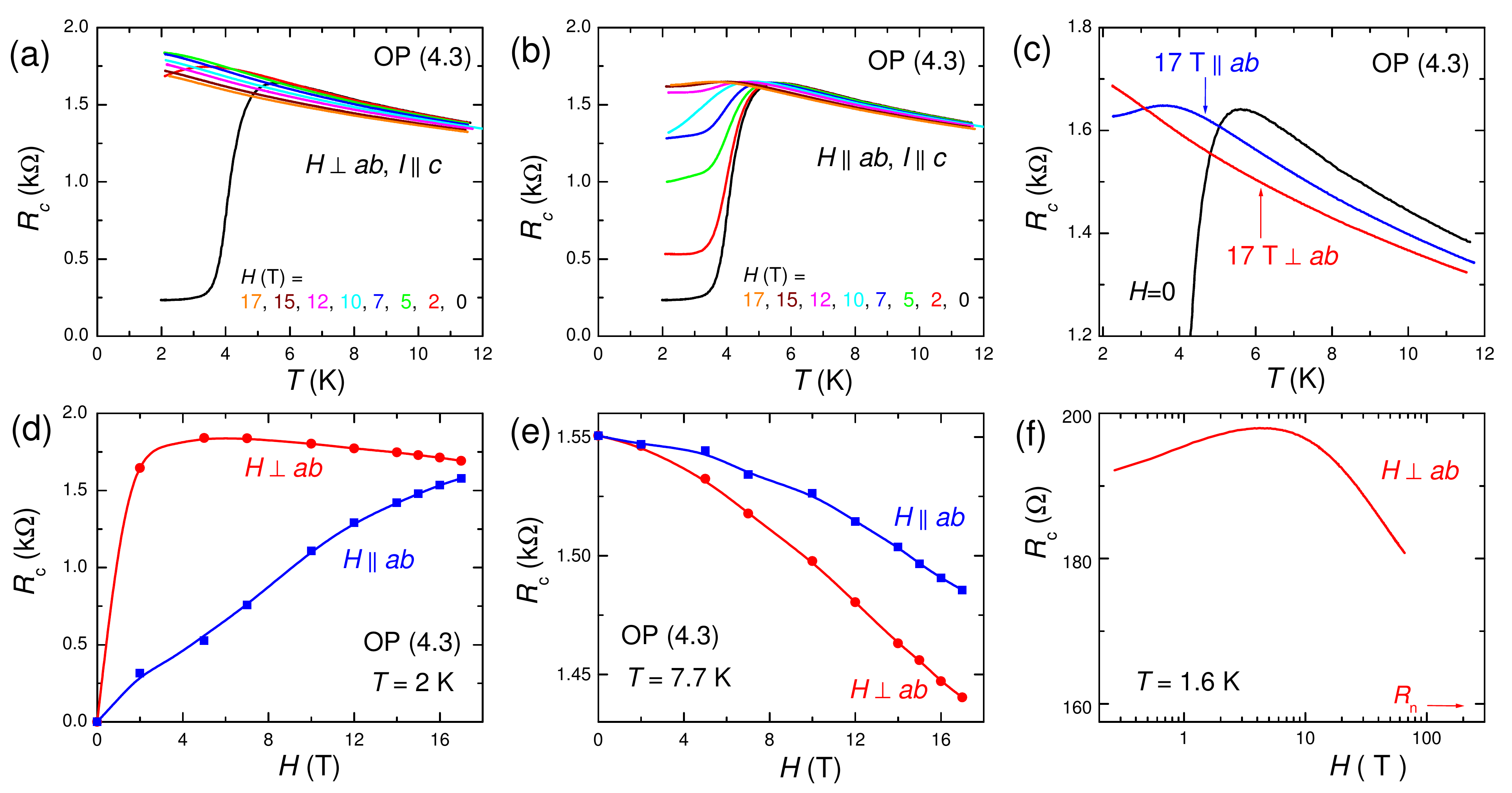}
\caption{(Color online) Temperature dependencies of the $c$-axis
resistance for (a) $H\perp ab$ and (b) $H\parallel ab$. (c)
Comparison of the data from (a) and (b) for $H=0$ and 17 T. Panels
(d) and (e) show $c$-axis MR for the two field orientations (d)
below and (e) above $T_c$. It is seen the normal state negative MR
is present for both field orientations. (f) $R_c(H^{\perp})$
measured up to 65 $T$ (data from Ref. \cite{Jacobs2012}). It is
seen that there is both a positive MR at low fields due to
suppression of the supercurrent and a negative MR at high fields
due to suppression of the PG. }
    \label{fig:fig3}
\end{figure*}

\subsection{B. Out-of-plane magnetoresistance}

Figure \ref{fig:fig3} (a) and (b) show temperature dependencies of
the $c$-axis resistance $R_c$ at different magnetic fields (a)
perpendicular and (b) parallel to the $ab$ planes.
Irrespective of field orientation, there are both positive and
negative contributions to $c$-axis MR. Fig. \ref{fig:fig3} (c)
represents a detailed comparison of $R_{c}(T)$ at $H=0$ and at
$H=17$ T for the two field orientations. It is seen that in the
normal state there is a significant negative $c$-axis MR for both
field orientations. It is largest for $H\perp ab$ and reaches
almost $10 \%$ in 17 T field.

A positive MR appears only in the superconducting state
$T<T_c(H)$. It is due to suppression of the interlayer Josephson
current with respect to the bias current
\cite{Morozov2000,Katterwe2008}. At $H
\parallel ab$ there is a profound Josephson flux-flow phenomenon
due to easy sliding of Josephson vortices along the $ab$-planes
\cite{Katterwe2009,Katterwe2010}. This also leads to a positive MR
with a peak at $H$ strictly parallel to the $ab$-planes
\cite{Motzkau2013}. The negative $c$-axis MR persists both in the
superconducting \cite{MR} and the normal states and is attributed
to field suppression of either the superconducting gap $\Delta$
\cite{MR} or the pseudogap $\Delta_{PG}$ \cite{KElbaum2005}. For
high-$T_c$ Bi-2212 \cite{KrTemp,SecondOrder} and Bi-2223
\cite{Suzuki_Bi2223} cuprates the corresponding energies ($\Delta
\sim 30-50$ meV, $\Delta_{PG} \sim 30-70$ meV) and fields
($H_{c2}\sim 100-200$ T, $H^*\sim 200-300$ T) are similar
\cite{MR,KElbaum2005} and separation of the two contributions is
difficult. However, in the studied low-$T_c$ superconductor the
separation becomes trivial because, as shown in Ref.
\cite{Jacobs2012}, all PG characteristics remain similar to
high-$T_c$ materials, but all superconducting characteristics
scale down with $T_c$ \cite{Ideta_ARPES}, leading to a large
disparity of superconducting and PG characteristics.

Figs. \ref{fig:fig3} (d,e) show $c$-axis MR for different field
orientations and temperatures (d) below and (e) above $T_c$. It is
seen that the negative MR persists at $H > H_{c2}^{\perp} \sim 10$
T and at $T>T_c$ and is due to field suppression of the PG
\cite{KElbaum2005,Jacobs2012}. Fig. \ref{fig:fig3} (f) shows
pulsed field measurements of $R_c(H^{\perp})$ at $T=1.6$ K up to
65 T for a slightly underdoped crystal from the same batch (data
from Ref. \cite{Jacobs2012}). It is seen that at high fields
$R_c(H^{\perp})$ is approximately linear in the semi-logarithmic
scale. An extrapolation to the normal resistance $R_n \sim 160~
\Omega$ yields the PG closing field $H^* \sim 300$ T. It
corresponds to the Zeeman energy $g\mu_B H^* \sim 35$ meV $\simeq
\Delta_{PG}$ \cite{Jacobs2012}.

\subsection{C. Angular magnetoresistance at $T<T_c$}

\begin{figure*}[t]
    \includegraphics[width=\textwidth]{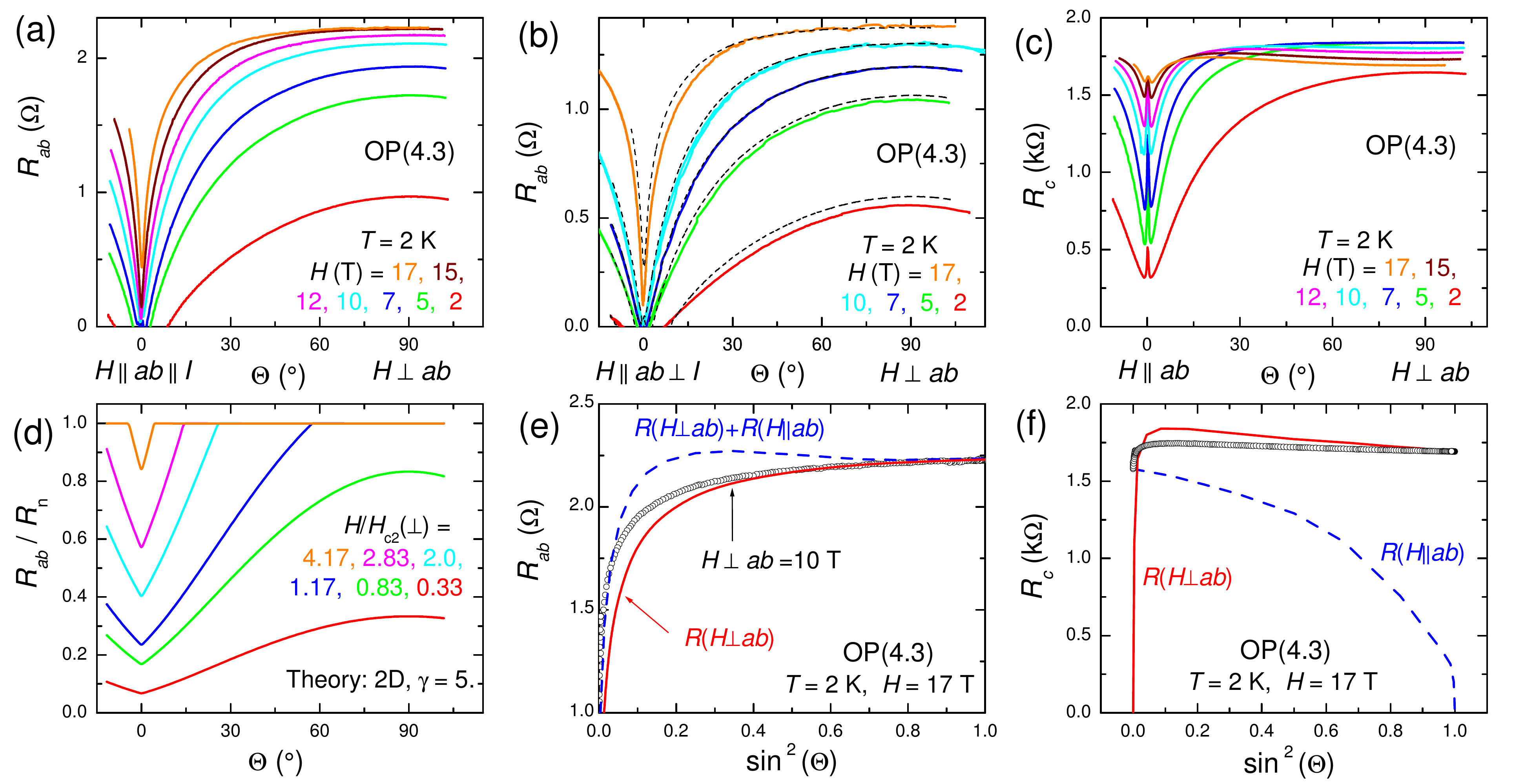}
\caption{(Color online) (a) and (b) Angular dependencies of
in-plane resistances for rotation around two orthogonal axes in
the $ab$-plane. Dashed lines in (b) represent scaled data from (a)
\cite{Note1}.
(c) Angular dependence of the $c$-axis resistance. The peak at
$\Theta=0^{\circ}$ is due to onset of Josephson flux-flow.
(d) Theoretical angular dependencies of flux-flow resistances for
a 2D model with an anisotropy $\gamma=5$. Note that the cusp at
$\Theta=0^{\circ}$ becomes sharper at $H>H_{c2}^{\perp}$ because
superconductivity survives only in a narrow range of angles around
$\Theta=0^{\circ}$. A similar narrowing is seen in panels (a-c).
Panels (e) and (f) represent comparison of (e) in-plane and (f)
out-of-plane angular MR (symbols) with resistances at the
corresponding perpendicular $H^{\perp}=H\sin(\Theta)$ (solid
lines) and parallel $H^{\parallel}=H\cos(\Theta)$ field components
at $T=2$ K and $H=17$ T. }
    \label{fig:fig4}
\end{figure*}

Angular dependence of the upper critical field $H_{c2}(\Theta)$ is
given by the following equations:

\begin{equation}\label{Hc23D}
    \left(\frac{H_{c2}(\Theta) \sin\Theta}{H^{\perp}_{c2}}\right)^2+ \left(\frac{H_{c2}(\Theta) \cos\Theta}{H^{\parallel}_{c2}}\right)^2=1,
\end{equation}
for a three-dimensional (3D) superconductor and

\begin{equation}\label{Hc22D}
    \left|\frac{H_{c2}(\Theta) \sin\Theta}{H^{\perp}_{c2}}\right|+   \left(\frac{H_{c2}(\Theta) \cos\Theta}{H^{\parallel}_{c2}}\right)^2=1,
\end{equation}
for the 2D case. In the simplest case of an isotropic
superconductor the flux-flow resistivity can be approximately
estimated from the Bardeen-Stephen model \cite{Stephen},

\begin{equation}\label{RvsAngle}
    R(\Theta)=R_n \frac{H}{H_{c2}(\Theta)}.
\end{equation}
It connects the angular MR $R(\Theta)$ with $H_{c2}(\Theta)$. The
main qualitative difference between 3D and 2D cases is that
$R(\Theta=0^{\circ})$ has a smooth minimum in the 3D case and a
sharp cusp-like dip in the 2D case \cite{Naughton1988}.

Figures \ref{fig:fig4} (a) and (b) show angular dependencies of
the in-plane resistance $R_{ab}(\Theta)$ at $T=2$ K measured upon
rotation around two orthogonal axes in the $ab$-plane (a)
perpendicular and (b) parallel to the current. In both cases
$\Theta=90^{\circ}$ corresponds to $H\perp ab$, $H\perp I$. But
$\Theta=0^{\circ}$, corresponds to either (a) the Lorentz
force-free configuration $H\parallel I$, or (b) to the case $H
\perp I$ when the Lorentz force is acting on Josephson vortices in
the direction perpendicular to layers. Dashed lines in (b)
represent properly scaled data from panel (a) \cite{Note1}. It is
seen that the behavior in both cases is very similar. Therefore,
at $H\parallel ab$ the flux-flow contribution to $R_{ab}$ is small
either due to zero Lorentz force or a strong intrinsic pinning in
the layered superconductor \cite{Tachiki,Kwok1991,NbCu1996}, which
prevents motion of Josephson vortices across the planes.

Fig. \ref{fig:fig4} (c) shows angular dependencies of the $c$-axis
resistance. Apart from the dip at $\Theta\sim 0^{\circ}$ due to
the anisotropy of $H_{c2}$, the $R_{c}(\Theta)$ has an additional
sharp maximum at $\Theta = 0^{\circ}$ due to onset of the
Josephson flux-flow phenomenon \cite{Motzkau2013}. In this case
the Lorentz force is directed along the $ab$-planes and easily
drags Josephson vortices with low pinning and viscosity
\cite{Katterwe2010}.
The shape of $R_c(\Theta)$ at large angles is visibly affected by
the negative normal state MR, which causes a shallow minimum of
$R_c(\Theta)$ at $\Theta = 90{^\circ}$ at large fields.

From Figs. \ref{fig:fig4} (a-c) it is seen that $R(\Theta)$
exhibits a cusp at $\Theta=0^{\circ}$, indicating the 2D-nature of
superconductivity in CuO planes. The cusp becomes narrower and
sharper with increasing field. This is in a qualitative agreement
with calculations for the 2D model using
Eqs.(\ref{Hc22D},\ref{RvsAngle}), shown in Fig. \ref{fig:fig4}
(d). The sharpening of the cusp at $\Theta=0^{\circ}$ occurs when
the field becomes larger than $H_{c2}^{\perp}$. In this case the
sample is in the normal state with a flat $R_{ab}(\Theta)= R_n$
for angles $\Theta \sim 90^{\circ}$ at which $H_{c2}(\Theta)<H$.
As the field approaches $H_{c2}^{\parallel}$, superconductivity
survives only in a narrow range of angles $\Theta \sim 0^{\circ}$.
Therefore, a significant narrowing of the cusp at $H=17$ T in Fig.
\ref{fig:fig4} (a-c) indicates that $H_{c2}^{\parallel}$ is close
to 17 T.

The anisotropy of $ H_{c2}$ can be analyzed from comparison of
angular-dependent $R(\Theta)$ with MR at the corresponding
parallel $R(H^{\parallel}=H\cos(\Theta))$ and perpendicular
$R(H^{\perp}=H\sin(\Theta))$ field orientations. If one of the
field components is smaller than the corresponding $H_{c2}$,
adding of an orthogonal component will contribute to suppression
of superconductivity. But if the field component is larger than
$H_{c2}$, than an extra field component will not give a
significant contribution to MR. In Figs. \ref{fig:fig4} (e) and
(f) we perform such the comparison at $T=2$ K. Black symbols in
Figs. \ref{fig:fig4} (e) represent $R_{ab}(\Theta)$ at $H=17$ T
from Fig. \ref{fig:fig4} (a) as a function of $\sin^2(\Theta)$.
The solid red line represents the MR in solely the perpendicular
field component $R_{ab}[H^{\perp}=H\sin(\Theta)]$. The dashed blue
line represents a sum of resistances in the corresponding
perpendicular and parallel field components
$R_{ab}[H^{\perp}=H\sin(\Theta)]+R_{ab}[H^{\parallel}=H\cos(\Theta)]$,
shown in Fig. \ref{fig:fig2} (e). It is seen that at
$\sin^2(\Theta) \gtrsim 0.35$ the angular MR is determined almost
entirely by $H^{\perp}$ and an additional $H^{\parallel}$ does not
contribute significantly to MR. This angle corresponds to
$H^{\perp} = H\sin(\Theta) > H_{c2}^{\perp} \simeq 10$ T, as
indicated by a vertical arrow in Fig. \ref{fig:fig4} (e). At
larger angles superconductivity is already suppressed because
$H^{\perp}> H^{\perp}_{c2}$ and MR becomes insensitive to an
additional parallel field component. Such the analysis confirms
that $H^{\perp}_{c2} \simeq 10$ T. At smaller angles $H^{\perp}<
H^{\perp}_{c2}$ and $H^{\parallel}$ does contribute to MR,
although not additively.

Fig. \ref{fig:fig4} (f) represents a similar comparison for the
out-of-plane resistance. Solid and dashed lines represent the MR
solely in perpendicular and parallel fields from Fig.
\ref{fig:fig3} (d). Apparently, $R_c(\Theta)$ is not determined by
a single field component. The most pronounced feature of
$R_c[\sin^2(\Theta)]$ is a rapid drop at $\sin(\Theta)\rightarrow
0$, which reflects the corresponding behavior of $R_c(H^{\perp})$.
Therefore, the crystal still maintains some superconductivity at
$H^{\parallel}=17$ T, but it is rapidly suppressed by a small
additional $H^{\perp}$ component upon a slight rotation of the
crystal. Consequently, $H^{\parallel}_{c2}(2{\text K})$ is
slightly larger than 17 T. On the other hand, since
$H^{\perp}_{c2}<17$ T, there is no similar drop at
$\sin(\Theta)=1$.

\begin{figure*}[t]
    \includegraphics[width=\textwidth]{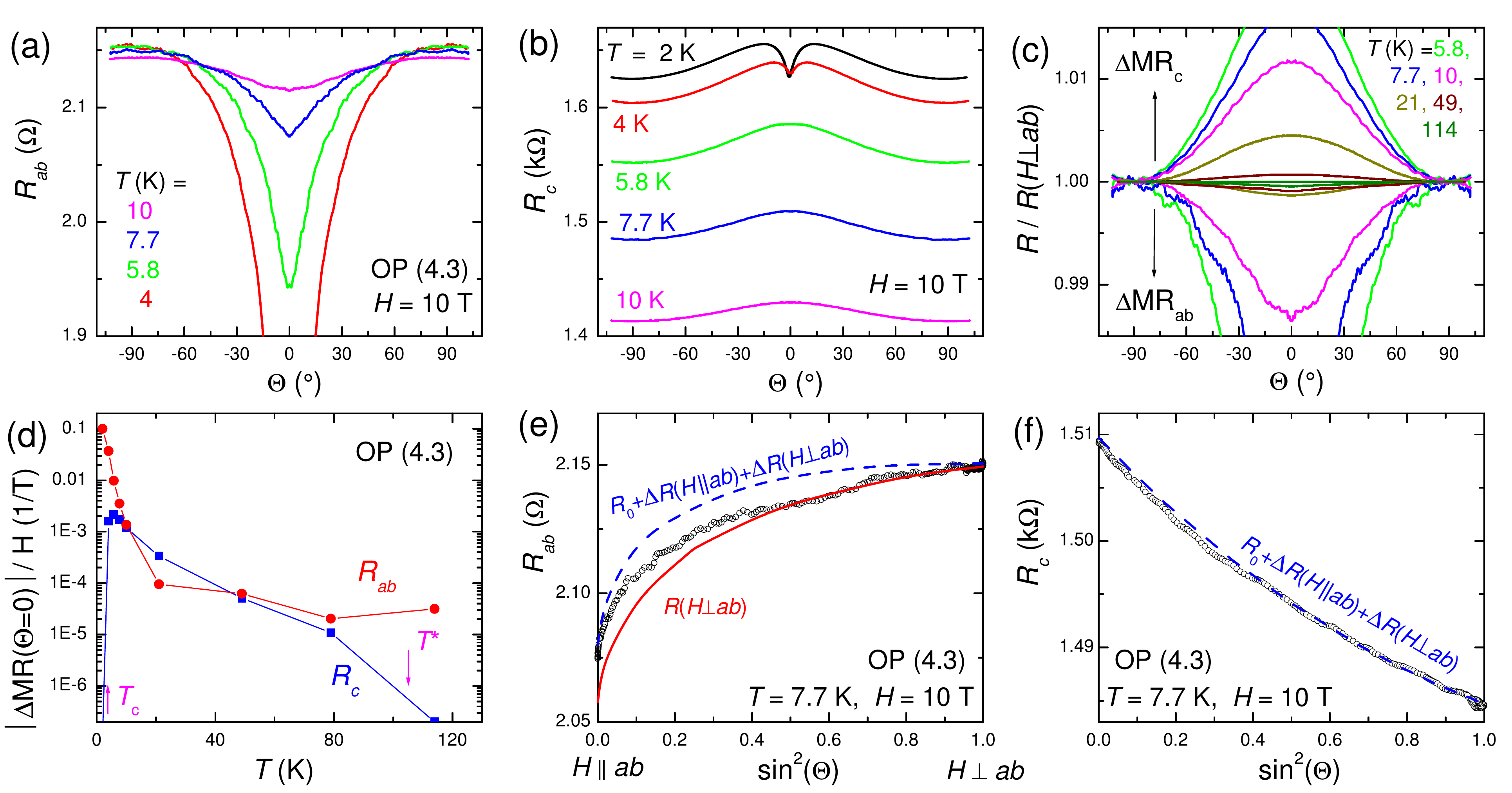}
\caption{(Color online) Angular dependence of (a) in-plane and (b)
$c$-axis resistances at $H=10$ T for different temperatures. The
cusp at $\Theta=0^{\circ}$ is vanishing at $T\gtrsim 2T_c$ for
$R_{ab}$ and at $T>T_c$ for $R_c$. (c) Angular dependent MR above
$T_c$, normalized by $R(\Theta=90^{\circ})$. At $T\geq 10$ K the
MR is varying in a smooth 3D-manner. (d) Temperature dependencies
of absolute values of angular MR amplitudes
normalized by the field (in the semi-logarithmic scale). Note that
the in-plane MR decays at a scale $T\sim T_c$ and the out-of plane
MR at the PG temperature $T^* \simeq 110$ K. Panels (e) and (f)
represent comparison of (e) in-plane and (f) out-of-plane angular
MR at $T=7.7$ K $>T_c$ with the MR at the corresponding parallel
and perpendicular field components. The in-plane angular MR at not
too small angles is dominated by the perpendicular field component
(solid line). The $c$-axis angular MR is given by the additive
contribution of the two field components (dashed lines).
}
    \label{fig:fig5}
\end{figure*}

\section{Fluctuation magnetoresistance}

From comparison of Figs. \ref{fig:fig4} (a), (b) and (d) it is
clear that Eqs. (\ref{Hc22D}) and (\ref{RvsAngle}) only explain
the narrowing of the cusp, but do not fit the $R(\Theta)$ data.
This demonstrates inappropriateness of Eq. (\ref{RvsAngle}) for
layered superconductors because it does not take into
consideration transformation of the vortex structure, the pinning
strength and the Lorentz force upon rotation of the crystal.
Furthermore, Eq. (\ref{RvsAngle}) assumes that the resistance
always reaches the normal state value $R_n$ at $H=H_{c2}$ and thus
neglects the remaining fluctuation para-conductivity at $H>H_{c2}$
\cite{Varlamov}. As discussed above, $R_{ab}(\Theta=0^{\circ})$
should have minimal flux-flow contribution either due to zero
Lorentz force, or presence of a strong intrinsic pinning.
Consequently, the dip in resistance at $\Theta=0^{\circ}$ in Fig.
\ref{fig:fig4} (a) and the major part of the resistive transition
$0<R<R_n$ at $H\parallel ab
\parallel I$ in Fig. \ref{fig:fig2} (b) are due to fluctuation
conductivity, rather than flux-flow. Without flux-flow, $H_{c2}$
would correspond to the onset of resistivity $R\sim 0$, rather
than $R=R_n$. This has been demonstrated by simultaneous tunneling
and transport measurements for conventional superconductors
\cite{MR}. Without exact knowledge of the flux-flow contribution
it is impossible to confidently extract $H_{c2}$ from $R(T,H)$
data at $T<T_c$. The lack of criteria for $R(H=H_{c2})$ obscures
estimation of $H_{c2}$ \cite{Vedeneev1999}. Therefore, in the
remaining part of the manuscript we will focus on the analysis of
fluctuation part of MR at $T>T_c$. As we will demonstrate, such
data do not suffer from ambiguity associated with flux-flow
phenomenon and facilitate confident extraction of $H_{c2}$.

\subsection{A. Angular magnetoresistance at $T>T_c$}

Figure \ref{fig:fig5} (a) shows angular dependencies of the
in-plane resistance at $H=10$ T and at different $T$ close and
above $T_c\simeq 4.3$ K. Here $\Theta =0^{\circ}$ corresponds to
zero Lorentz force configuration $H\parallel ab \parallel I$. It
is seen that the cusp at $\Theta =0^{\circ}$, characteristic for
the 2D superconducting state, is rapidly diminishing with
increasing $T>T_c$. It disappears at $\sim 2 T_c$. At $T\gtrsim
10$ K it turns into a shallow minimum, which persists to $T\gg
T_c$ and represents the anisotropy of the positive orbital MR in
the normal state.

Fig. \ref{fig:fig5} (b) shows angular dependencies of the $c$-axis
resistance below and above $T_c$. Here, measurements were
performed at
bias above the Josephson flux-flow branch in the $I$-$V$ so that
the Josephson flux-flow peak in $R_c(\Theta=0^{\circ})$ does not
occur \cite{Motzkau2013}. Above $T_c$ the cusp in
$R_{c}(\Theta)=0^{\circ}$ completely disappears and only a shallow
maximum at $\Theta = 0^{\circ}$ remains, which indicates a small
angular anisotropy of the normal state MR, as seen from Fig.
\ref{fig:fig3} (e). In Fig. \ref{fig:fig5} (c) we show angular
dependencies of in-plane and $c$-axis resistances, normalized by
the corresponding values at $\Theta=90^{\circ}$. One can see a
shallow 3D behavior in the normal state.

\begin{figure*}[t]
    \includegraphics[width=\textwidth]{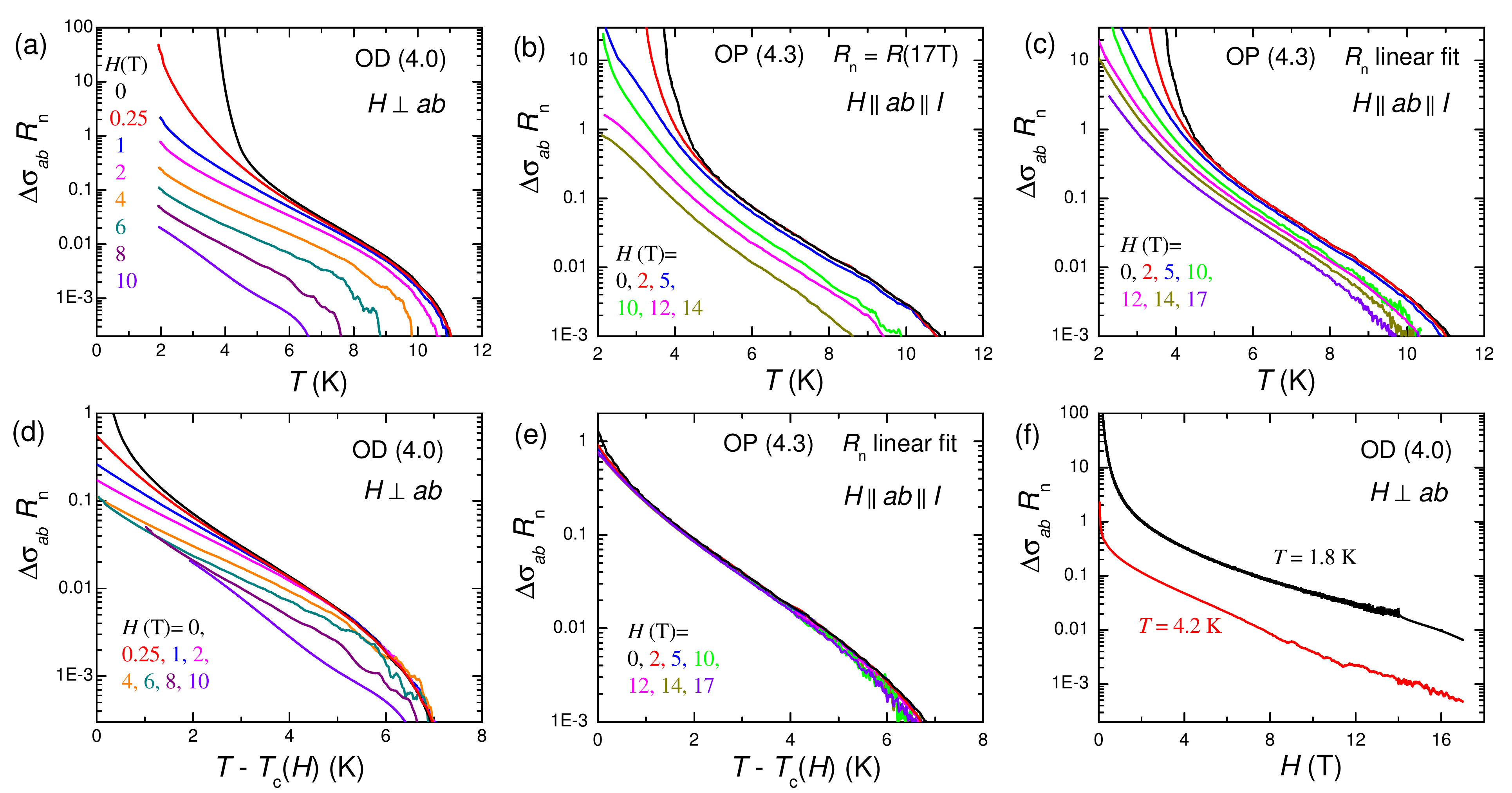}
\caption{(Color online) Fluctuation part of the in-plane
conductivity $\Delta \sigma_{ab}(T) = 1/R_{ab}(T)-1/R_n(T)$,
normalized by the normal state resistance, for fields (a)
perpendicular and (b,c) parallel to the $ab$-plane. Curves in
panels (b) and (c) were obtained from the same data using
different $R_n(T)$: (b) $R_n=R_{ab}(H=17$ T), (c) linear
extrapolation from high $T$, shown by the dashed line in Fig.
\ref{fig:fig2} (b). It is seen that fluctuation para-conductivity
decays approximately exponentially with increasing $T$ with an
almost field independent slope. Panels (d) and (e) show data from
(a) and (c), respectively, shifted by $T_c(H)$. (f) Field
dependence of para-conductivity $\Delta \sigma_{ab}(H^{\perp})$ at
different $T$. An approximately exponential decay is seen.}
    \label{fig:fig6}
\end{figure*}

In Fig. \ref{fig:fig5} (d) we show absolute values of the angular
MR amplitude $\mid
\Delta$MR$(\Theta=0^{\circ})=R(0^{\circ})/R(90^{\circ})-1 \mid$,
normalized by the magnetic field, for the in-plane and the
$c$-axis resistances.
The in-plane $\Delta$MR$_{ab}$ (circles) is large in the
superconducting state and remains significant in the fluctuation
region at $T_c<T\lesssim 2 T_c$ when the cusp in
$R_{ab}(\Theta=0^{\circ})$ is observed, see Fig. \ref{fig:fig5}
(a). With increasing $T$, $\mid \Delta$MR$_{ab} \mid$ rapidly
decreases. At $T> 20$ K it flattens off. The remaining weakly
$T$-dependent value represents the anisotropy of the positive
in-plane MR in the normal state, presumably of the orbital origin.
The out-of-plane $\mid\Delta$MR$_{c}\mid$ (squares) decreases
almost exponentially with increasing temperature in a wide
$T$-range above $T_c$. It becomes hardly detectable above the
pseudogap opening temperature $T^*\simeq 110$ K, while the
in-plane $\Delta$MR$_{ab}$ still remains recognizable.

A different behavior of in-plane and out-of-plane MR can be also
seen from comparison of individual and combined contributions of
the two field components. Symbols in Figs. \ref{fig:fig5} (e) and
(f) show angular dependent (e) in-plane and (f) $c$-axis MR at
$T=7.7$ K $>T_c$ as a function of $\sin^2 (\Theta)$. Dashed blue
lines represent additive contributions from the two field
components, $R=R_0+\Delta R(H^{\perp}) +\Delta R(H^{\parallel})$,
where $R_0=R(H=0)$, and $\Delta R(H^{\perp})$ and $\Delta R
(H^{\parallel})$ are the corresponding MR solely in perpendicular
and parallel fields, shown in Figs. \ref{fig:fig2} (f) and
\ref{fig:fig3} (e). It is seen that the $c$-axis MR is well
described by the simple additive contribution of the two field
components, while the in-plane does not. This reflects different
mechanisms of in-plane and out-of-plane magnetoresistances. The
negative $c$-axis MR is due to field suppression of the pseudogap.
The applied field is much smaller than the PG closing field
$H^*\sim 300$ T \cite{Jacobs2012}. Therefore, the $c$-axis MR is
far from saturation and is approximately linear in field, leading
to additive, independent from each other, contribution from the
two field components.

The positive in-plane MR at $T_c<T\lesssim 2 T_c$ is mostly due to
suppression of superconducting fluctuations with the
characteristic field $H^{\perp}_{c2} \sim 10$ T, which is in the
range of applied fields. This leads to saturation of MR and to
non-additive contribution of the two field components. Unlike the
normal state angular MR, which has a 3D character, as shown in
Fig. \ref{fig:fig5} (c), superconducting fluctuations at
$T_c<T\lesssim 2T_c$ remain quasi-2D, as seen from the cusp in
$R_{ab}(\Theta=0)$ in Fig. \ref{fig:fig5} (b). The solid line in
Fig. \ref{fig:fig5} (e) indicates that at not too small angles the
in-plane MR is determined by the $c$-axis field component.

\subsection{B. Fluctuation conductivity}

Fluctuation para-conductivity is seen as a tail of the in-plane
resistive transitions from Figs. \ref{fig:fig2} (a) and (b) at
$T_c<T\lesssim 10$ K, in the same range where the cusp is seen in
the angular MR, Fig. \ref{fig:fig5} (a). Figures \ref{fig:fig6}
(a-c) represent normalized excess conductivities $\Delta
\sigma_{ab}(T) = 1/R_{ab}(T)-1/R_n(T)$, in perpendicular and
parallel magnetic fields.
Here we used different approximations for $R_n$: (a)
$R_n^{\perp}(T)=R_{ab}(T,H^{\perp}=14~{\text T})$, (b)
$R_n^{\parallel}(T)=R_{ab}(T,H^{\parallel}=17~{\text T})$, and (c)
a linear extrapolation from high $T$, shown by the dashed line in
Fig. \ref{fig:fig2} (b).

It is seen that for both field orientations the fluctuation
conductivity at $T>T_c$ decreases approximately linearly in the
semi-logarithmic scale with almost field-independent slopes. This
implies
\begin{equation}\label{Sigma_TH}
    \Delta\sigma_{ab}(T,H) \propto \exp[-a(T-T_c(H))],
\end{equation}
where $a$ is some constant. A similar exponential decay has been
reported for other cuprates \cite{MR,Alloul2011,Luo2014}. Even
though such an exponential decay does not follow explicitly from
theoretical analysis of fluctuation conductivity
\cite{Varlamov,Finkelstein2012}, it allows an unambiguous
determination of the characteristic temperature scale $T_c(H)$
from the relative shift of the curves along the $T$-axis with
respect to the known $T_c(H=0)$. Since the $\Delta \sigma_{ab}(T)$
curves in Figs. \ref{fig:fig6} (a-c) remain almost parallel at
different $H$, such determination of $T_c(H)$ does not suffer from
widening of the resistive transition, as in the flux-flow case at
$T<T_c$ in Fig. \ref{fig:fig2} (a). Therefore, thus obtained
$T_c(H)$ has the same degree of certainty as $T_c(H=0)$.

\begin{figure*}[t]
    \includegraphics[width=\textwidth]{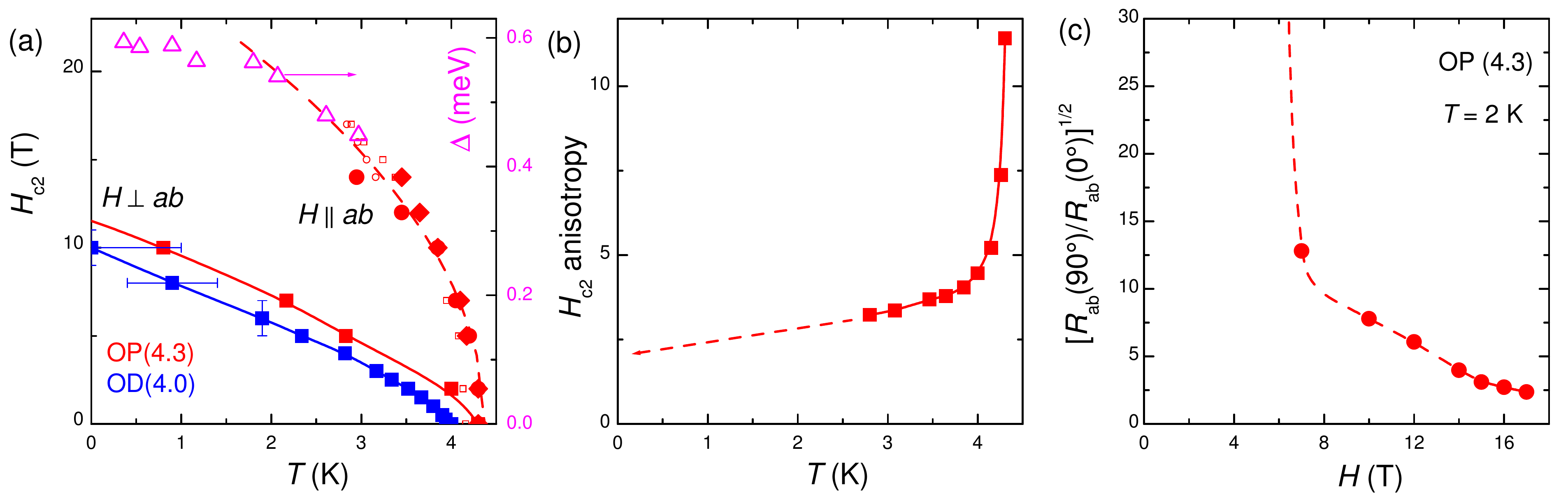}
    \caption{(Color online)
(a) The upper critical field perpendicular (filled squares) and
parallel to layers (filled circles and rhombuses) obtained from
the scaling analysis of fluctuation conductivity according to Eq.
(\ref{Sigma_TH}) at $T>T_c(H)$.
For comparison we also show middle points $H_{50\%}(T)$ of
in-plane (small open circles) and out-of-plane (small open
squares) resistive transitions in parallel field. The dashed line
represents the $\sqrt{T_c-T}$ dependence. Open triangles (right
axis) show $T$-dependence of the superconducting gap, obtained by
intrinsic tunneling spectroscopy \cite{Jacobs2012}. (b)
$T$-dependence of the anisotropy of the upper critical field
$\gamma_H=H_{c2}^{\parallel}/H_{c2}^{\perp}$. It is seen that at
low temperature it saturates at a small value $\gamma_H(0) \sim
2$. (c) Angular anisotropy of the in-plane resistance
$R_{ab}(90^{\circ})/R_{ab}(0^{\circ})$ at $T=2$ K as a function of
magnetic field. The anisotropy rapidly decreases with increasing
$H$ as soon as the field approaches the paramagnetic limit.}
    \label{fig:fig7}
\end{figure*}

Equation (\ref{Sigma_TH}) suggests that $\Delta \sigma (T,H)$
curves could be collapsed in one by shifting them along the
$T$-axis by $T_c(H)$. In Figs. \ref{fig:fig6} (d) and (e) we show
such an attempt for the data from Figs. \ref{fig:fig6} (a) and
(c), respectively. Even though the scaling is not always perfect,
the shift parameter $T_c(H)$ is determined unambiguously because:
(i) The shift for the curve at $H=0$ is fixed by $T_c(0)$. (ii)
The curves from low to intermediate fields do collapse at high
enough $T$. (iii) When the curves do not collapse, we required
that fluctuation conductivity for a given $T-T_c(H)$ should be
decreasing with increasing $H$ because superconductivity is
suppressed by magnetic field. This means that the $\Delta \sigma
(T-T_c(H))$ curves at higher $H$ should always lie lower and
should not cross the curves at smaller $H$. In Fig. \ref{fig:fig6}
(d) the curve $\Delta \sigma (T, H=10 ~{\text T})$ was not shifted
at all, implying that $T_c(H=10 ~{\text T}) =0$, which is
consistent with our previous estimation of
$H^{\perp}_{c2}(T=0)\simeq 10$ T.

Fig. \ref{fig:fig6} (f) represents a semi-logarithmic plot of
$\Delta \sigma_{ab} R_n$ vs $H\perp ab$ for the OD(4.0) sample at
$T=1.8$ K and slightly above $T_c$ at $T=4.2$ K. It is seen that
$\Delta \sigma_{ab} (H)$ decays almost exponentially also as a
function of field at constant $T$. In this case the relative shift
along the horizontal axis provides the characteristic magnetic
field scale for suppression of superconductivity $\sim H_{c2}$.
Assuming that $H_{c2}=0$ at $T=4.2~{\text K} \sim T_c$, we
estimate from the relative shift of the two curves that
$H^{\perp}_{c2}(T=2~K)\simeq 6$ T. This is consistent with
$T_c(H^{\perp}=6~{\text T})\simeq 2~{\text K}$, estimated from
$\Delta \sigma_{ab} (T)$ scaling in Fig. \ref{fig:fig6} (d). Thus,
from the analysis of fluctuation conductivity we obtain a
confident estimation of $T_c(H)$ or equivalently $H_{c2}(T)$.

\subsection{C. The upper critical field}

Figure \ref{fig:fig7} (a) contains the main result of this work:
$T$-dependencies of $H_{c2}$ obtained from the analysis of
fluctuation conductivity, Eq.(\ref{Sigma_TH}), at $T>T_c(H)$
(filled symbols). Filled blue and red squares represent
$H^{\perp}_{c2}(T)$ for OD(4.0) and OP(4.3) crystals,
respectively. Horizontal and vertical error bars correspond to the
accuracy of scaling of $\Delta \sigma (T,H)$ curves according to
Eq. (\ref{Sigma_TH}), as seen in Figs. \ref{fig:fig6} (d) and (f).

Estimation of $H^{\parallel}_{c2}$ at low $T$ is complicated by
the lack of confident knowledge of $R_n(T)$. In Fig.
\ref{fig:fig6} (b) and (c) we used two different approximations of
$R_n(H^{\parallel})$. Filled circles and rhombuses represent
$H^{\parallel}_{c2}(T)$ for the OP(4.3) crystal, obtained from the
scaling of data in Fig. \ref{fig:fig6} (b) and Figs.
\ref{fig:fig6} (c, f), respectively. Up to $H^{\parallel}=10$ T
both approximations of $R_n$ give the same
$H^{\parallel}_{c2}(T)$. Therefore, those values are confident.
However, at $H>12$ T results start to depend on the choice of
$R_n(T)$. Unfortunately, none of the two approximations 
is good enough at $T\rightarrow 0$. Qualitatively,
$R_n=R_{ab}(H^{\parallel}=17~{\text T})$ tends to underestimate
$H_{c2}^{\parallel}$ because it assumes $H_{c2}^{\parallel}
(T=0)=17$ T. The linear extrapolation of $R_n(T>T_c)$ tends to
overestimate $H_{c2}^{\parallel} (T=0)$ because it assumes that
$R(H=H_{c2})=R_n$.
However, without the flux-flow phenomenon $R(H=H_{c2})\simeq 0$
\cite{MR}. This is what we expect for our Lorentz force free data
at $H\parallel ab$. In absence of a better way to define
$H_{c2}^{\parallel}$ at low $T$, in Fig. \ref{fig:fig7} (a) we
also show fields $H_{50\%}(T)$ at which middle points of resistive
transitions occurs for in-plane (open circles) and $c$-axis (open
squares) resistances. Those points fall inbetween the
underestimating (solid circles) and overestimating (rhombuses)
analysis of fluctuation conductivity. Therefore, they provide a
reasonable estimate of $H_{c2}^{\parallel}$ at lower $T$.

From Fig. \ref{fig:fig7} (a) it is seen that $H^{\perp}_{c2}(T)$
and $H^{\parallel}_{c2}(T)$ are qualitatively different. The
$H^{\perp}_{c2}(T)$ is almost linear in the whole $T$-range
$H^{\perp}_{c2}(T) \propto T_c-T$. Such a behavior is consistent
with a conventional orbital upper critical field,
\begin{equation}\label{Hc2_perp}
H^{\perp}_{c2}=\frac{\Phi_0}{2\pi\xi_{ab}^2},
\end{equation}
where $\Phi_0$ is the flux quantum and $\xi_{ab}$ is the in-plane
coherence length, $\xi_{ab}(0)=5.5 \pm 0.2$ nm.

The $H^{\parallel}_{c2}(T)$ is clearly non-linear. The dashed line
in Fig. \ref{fig:fig7} (a) demonstrates that $H^{\parallel}_{c2}
(T) \propto \sqrt{T_c-T}$. At the first glance, it resembles the
behavior of $H^{\parallel}_{c2} (T)$ in thin film multilayers
\cite{Tachiki,NbCu1996},
\begin{equation}\label{Hc2_ll}
H^{\parallel}_{c2}=\frac{\sqrt{3}\Phi_0}{\pi\xi_{ab}d},
\end{equation}
where $d$ is the thickness of superconducting layers. However, the
corresponding $d=9.3 \pm 0.5$ nm is much larger than the thickness
of CuO layers $\sim 0.2$ nm, as noted previously in Ref.
\cite{Vedeneev2006}, and is not connected to any geometrical
length scale of the sample. Consequently, there is no agreement
with Eq. (\ref{Hc2_ll}).

\subsection{D. The paramagnetic limit}

The upper limit of $H_\text{c2}$ is determined by Pauli
paramagnetism. The spin-singlet pairing is destroyed when the
Zeeman spin-split energy becomes comparable to the superconducting
energy gap $\Delta$. This gives \cite{SaintJames,Clogston}
\begin{equation}\label{ParamagnHc2}
    H_p=\frac{\sqrt{2}\Delta}{g \mu_B},
\end{equation}
where $g$ is the gyromagnetic ratio and $\mu_B$ is the Bohr
magneton. In case of negligible spin-orbit coupling $g\simeq 2$
this yields $\mathrm dH_\text{c2}/\mathrm dT(T=T_\text{c}) =
-2.25$\,T/K for d-wave superconductors \cite{ParamagneticLimit}.
Our values $H^{\perp}_\text{c2}/T_\text{c} \simeq 2.5$ T/K and
$\mid\mathrm dH^{\perp}_\text{c2}/\mathrm dT\mid (T=T_\text{c})
\simeq 5$ T/K and especially $H^{\parallel}_\text{c2}/T_\text{c}
\simeq 5$ T/K and $\mid\mathrm dH^{\parallel}_\text{c2}/\mathrm
dT\mid (T=T_\text{c})>40$ T/K clearly exceed this limit. Most
importantly, $H_p$ does not depend on orientation of the field.
Therefore, paramagnetically limited $H_{c2}$ should be
approximately isotropic, irrespective of the underlying effective
mass anisotropy.

According to Eq.(\ref{ParamagnHc2}), $H_p$ is determined solely by
$\Delta$. Open triangles in Fig. \ref{fig:fig7} (a) show
$\Delta(T)$-dependence measured by intrinsic tunneling
spectroscopy on a slightly underdoped crystal from the same batch
\cite{Jacobs2012}. It matches nicely $H_{c2}^{\parallel}(T)$.
Therefore, we conclude that the observed $H^{\parallel}_{c2} (T)
\propto \sqrt{T_c-T}$ dependence is not originating from the
geometrical confinement, Eq.(\ref{Hc2_ll}), but follows the
corresponding $\Delta(T)$ dependence of $H_p$ in
Eq.(\ref{ParamagnHc2}).

Fig. \ref{fig:fig7} (b) shows the anisotropy of the upper critical
field $\gamma_H=H^{\parallel}_\text{c2}/H^{\perp}_\text{c2}$.
Close to $T_c$ it diverges due to different $T$-dependencies of
the two fields. However, at $T\ll T_c$ it shows a tendency for
saturation at $\gamma_H(T\rightarrow 0) \sim 2$. Such a low
anisotropy of $H_{c2}$ is remarkable for the layered Bi-2201
compound with
$\gamma_m \sim 300$ \cite{Vinnikov}.

In Fig. \ref{fig:fig7} (c) we show magnetic field dependence of
the angular anisotropy
$[R_{ab}(90^{\circ})/R_{ab}(0^{\circ})]^{1/2}$ obtained from the
data in Fig. \ref{fig:fig4} (a). The anisotropy is large at low
fields, but rapidly decreases at $H > 7$ T when the paramagnetic
limitation starts to play a role. At high fields it tends to
saturate at $\sim 2$, consistent with $\gamma_H$ in Fig.
\ref{fig:fig7} (b). As mentioned above, paramagnetically limited
$H_{c2}$ should be isotropic. Therefore, a finite residual
anisotropy $\gamma_H(T\rightarrow 0) \sim 2$ indicates that only
$H_{c2}^{\parallel}$ is paramagnetically limited, while
$H_{c2}^{\perp}$ is still governed by orbital effects. Finally we
note that $\gamma_H<\gamma_m$ was reported for several
unconventional superconductors
\cite{Sekitani2004,Vedeneev2006,Zuo2000,Singleton2000}.
In particular, a nearly isotropic $H_{c2}$ was reported for the
(Ba,K)Fe$_2$As$_2$ pnictide \cite{Yuan2009} despite a quasi-2D
electronic structure. It is likely that all those observations
have the same origin.

\section{Conclusions}

To conclude, we presented a comprehensive analysis of both
in-plane and out-of-plane magnetoresistance in a layered cuprate
Bi$_{2.15}$Sr$_{1.9}$CuO$_{6+\delta}$ with a low $T_c\simeq 4$ K.
We have shown that the in-plane and the out-of-plane resistances
behave differently almost in all respects. The in-plane
magnetoresistance has two positive contributions. The positive
in-plane MR due to suppression of superconductivity (or
superconducting fluctuations) is dominant at $T\lesssim 2T_c$ and
magnetic fields $H_{c2} \lesssim 10$ T. It is clearly
distinguishable by its 2D cusp-like angular dependence. At $T
\gtrsim 2T_c$ the superconducting contribution vanishes and only a
weakly $T$-dependent positive MR, presumably of orbital origin,
remains. Such normal state in-plane MR has a smooth 3D-type
angular dependence. The $c$-axis MR at $T>T_c$ is dominated by a
negative MR caused by suppression of the pseudogap. It decays
rapidly upon approaching the PG opening temperature $T^* \simeq
110$ K $\gg T_c$ and at the PG closing field $H^*\sim 300$ T $\gg
H_{c2}$, and exhibits a smooth 3D-type angular dependence.
Different behavior of the in-plane and the out-of-plane MR
underlines different origins of superconductivity and the $c$-axis
pseudogap, which becomes particularly obvious from analysis of
low-$T_c$ cuprates \cite{Jacobs2012}.

The main focus of our work was on analysis of fluctuation
conductivity at $T>T_c$. We observed a universal, nearly
exponential, decay of in-plane para-conductivity as a function of
temperature and magnetic field and proposed a method for
extraction of $H_{c2}$ based on a new type of a scaling analysis
of the fluctuation para-conductivity. This way we obtained
confident values of $H_{c2}$, avoiding the complexity of flux-flow
phenomena at $T<T_c$. We observed that $H_{c2}^{\perp}$ is
following a linear $T$-dependence $H_{c2}^{\perp} \propto
1-T/T_c$, typical for $H_{c2}$ limited by orbital effects. On the
other hand, $H_{c2}^{\parallel}$ follows the $T$-dependence of the
superconducting gap with a characteristic $\propto \sqrt{1-T/T_c}$
dependence close to $T_c$. Our main result is observation of a
remarkably low anisotropy of the upper critical field
$\gamma_H(T\rightarrow 0) \simeq 2$, which is much smaller than
the effective mass anisotropy $\gamma_m \sim 300$. This
demonstrates that the anisotropy of $H_{c2}$ in unconventional
superconductors may have nothing to do with the anisotropy of the
electronic structure and the actual anisotropy of
superconductivity at zero field. The large discrepancy in
anisotropies serves instead as a robust evidence for
paramagnetically limited superconductivity.

\subsection{Acknowledgements}

Technical support from the Core Facility in Nanotechnology at
Stockholm University is gratefully acknowledged. We are grateful
to A. Rydh and M.V. Kartsovnik for assistance in experiment and
useful remarks.

\end{document}